\documentclass{pspum-l}
\usepackage{pictex,graphicx}

\newtheorem{theorem}{Theorem}[section]

\theoremstyle{definition}

\theoremstyle{remark}

\numberwithin{equation}{section}

\begin{document}

\newcommand{\Ai}{\mathop{\rm Ai}}
\newcommand{\T}{\overline{T}}

\title{Investigating the Spectral Geometry of a Soft Wall}


\author{J. D. Bouas}
\address{Department of Mathematics, Texas A\&M 
University, College Station, TX, 77843-3368, U.S.A.}
\curraddr{}
\email{jd.bouas@gmail.com \rm(Jeff Bouas)}
\thanks{Supported by National Science Foundation Grants Nos.\ 
PHY-0554849 and PHY-0968269.}

 \author{S. A. Fulling}                     
\email{fulling@math.tamu.edu \rm(Stephen Fulling)}

\author{F. D. Mera}
\email{fmera@ucmerced.edu \rm(Fernando Mera)}

\author{K. Thapa}
\email{thapakrish@tamu.edu \rm(Krishna Thapa)}

\author{C. S. Trendafilova}
\email{cyntrendafilova@gmail.com \rm(Cynthia Trendafilova)}

\author{J. Wagner}
\curraddr{Department of Physics and Astronomy,
University of California -- Riverside, 3401 Watkins Dr., 
Riverside, CA 92521 OK}
\email{jeffrey.wagner@ucr.edu \rm(Jef Wagner)}

\date{May 31, 2012}

\begin{abstract}
The idealized theory of quantum vacuum energy density is a beautiful 
application of the spectral theory of differential operators with 
boundary conditions, but its conclusions are physically 
unacceptable.  A more plausible model of a reflecting boundary 
 that stays within linear spectral theory  confines the waves by a 
steeply rising potential function, which can be taken as a power of 
one coordinate, $z^\alpha$.
 We report investigations of this model with considerable student 
involvement.
An exact analytical solution with some numerics for $\alpha=1$ and 
an asymptotic (semiclassical) analysis of a related problem 
 for $\alpha=2$ are presented. 
 \end{abstract} 

\maketitle

 \section{Introduction}

 The Casimir effect \cite{casimir,BmC,PMG,milton,BMM} 
 is an observable attraction 
between neutral electrical conductors.  Its mathematical charm is 
that, at least  for perfect conductors, it can be attributed to and 
calculated from the geometrical dependence of the energy of the 
quantized electromagnetic field in the region between the 
conductors.  As often  in quantum field theory, naive 
calculations give an infinite answer.   Subtraction of 
the zero-point energy of each field mode  renders the local 
energy density finite but leaves a nonintegrable singularity at the 
boundaries.  This divergence, because it can be regarded as 
permanently attached to the conductors, does not interfere with the 
calculation of the force of attraction between rigid bodies. 
 However, a more accurate representation of the physics near the 
boundary is needed \cite{barton1,jaffe} to understand situations where 
the boundary can deform, such as the celebrated case of an 
expandable sphere \cite{boyer}.
 Furthermore, the energy density (more completely, the stress tensor)
  of the field 
should act as a source of the gravitational field in general 
relativity, so  localized infinities within it are not physically 
plausible \cite{DC}.
 For more detailed background information see 
\cite{fulqfext05,fuliowa,EFKKLM,rect,kaplan}.

 It is universally agreed among physicists that the root of this 
problem is that the idealization of a ``perfect conductor'' is 
inapplicable to very-high-frequency modes of the quantized field.
 A full treatment of the physical problem, including  modeling of the 
charged particles inside the conductors, would take us out of the 
spectral theory of self-adjoint linear partial 
differential operators into difficult, nonlinear condensed-matter 
physics \cite{barton2}.  One might hope that an ad hoc cutoff of the 
contribution of high-frequency modes would preserve the qualitative 
essence of the physically correct solution, and indeed a simple 
exponential cutoff leads to both tractable calculations and 
physically plausible results \cite{rect,kaplan}, while placing 
the topic 
firmly within the study of the asymptotics of integral kernels 
(Green functions) associated with the operator concerned, in the 
grand tradition of spectral and geometrical asymptotics. 
 (Most of the issues of principle in vacuum energy are adequately 
addressed by studying a scalar field with Dirichlet boundary 
conditions instead of an electromagnetic field with 
 perfect-conductor boundary conditions, and our  discussion has now 
lapsed into that setting.)

 Unfortunately, a close examination of the stress tensor (in 
particular, energy density and pressure) predicted by the theory of 
\cite{rect} reveals that the expected energy-balance equation (sometimes 
called principle of virtual work \cite{barton3}),
 \begin{equation} \frac{\partial E}{\partial h} = - \int_{S_h} 
p\,,
 \label{enbalance}\end{equation}
 describing the change in total energy when a boundary acted upon 
by pressure~$p$ is moved a distance~$h$,
 is not satisfied \cite{fulqfext09}.
 The root of this problem appears to be that the degree to which a 
particular normal mode is affected by the cutoff  depends on~$h$, 
so that derivatives of the cutoff function contaminate fundamental 
relations like (\ref{enbalance}).  The exponential cutoff on 
frequency amounts, after analytic continuation, to a time-direction 
separation of the space-time points that are the argument variables 
of the integral kernel for the wave equation of the problem.
 Physically plausible results can be achieved for the various 
tensor components in various scenarios by choosing a specific
  space-time direction for the point separation in each case, but 
  such an ad hoc procedure cannot be regarded as logically 
  satisfactory for the long term.

Our present goal is to 
replace the reflecting boundary and the cutoff with a smooth 
potential rising to infinity.
The potential can be thought of as a static 
configuration  of another 
scalar field (presumably of very high mass).
  After some well understood 
renormalizations in the bulk, this model should yield a finite 
vacuum stress tensor without any cutoff.  As an internally 
consistent physical system without ad hoc modification, it is 
expected to satisfy the physically required energy-balance 
relation. Yet in the limit of a very steep wall it should approach 
in some sense the vacuum stress of the problem with Dirichlet 
boundary.
If its predictions  resemble those for 
 hard-wall calculations with certain point 
separations, those  (technically easier) ad hoc methods will be 
vindicated and can be used with confidence in other situations.


 Here we report progress on this problem achieved during the spring 
semester of 2010, while J.~Wagner held a Visiting Assistant 
Professorship in Mathematics at Texas A\&M University.
 He worked closely with Professor S.~Fulling and four research 
assistants supported by his NSF grant:  mathematics M.S. students 
J.~Bouas and F.~Mera and undergraduates (physics-mathematics double 
majors) C. Trendafilova and K.~Thapa.
 This brief period was notable for intensity of collaboration.

 \section{The model}

\begin{figure}
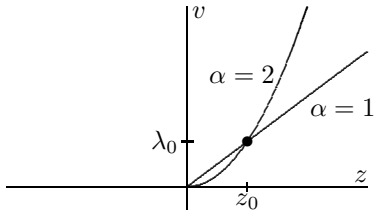

 \[\beginpicture
\setcoordinatesystem units <0.8cm,0.6cm>
 \putrule from -3 0 to 3 0
 \putrule from 0 -0.5 to 0 4
  \put{$z$} [rb] <0pt,2pt> at 3 0
   \put{$v$} [lt] <2pt,0pt> at  0 4
 \plot 0 0
       3 3 /
 \setquadratic
  \plot 0 0
        1 1
        2 4 /
 \put{$\alpha=1$} [lt] <1pt,-1pt> at 2 2
\put{$\alpha=2$} [rb] <-1pt,1pt> at 1.5 2.25
 \putrule from 1 -0.1 to 1 0.1
 \putrule from  -0.1 1 to 0.1 1                       
 \put{$z_0$} [t] <0pt,-1pt> at 1 -0.1
\put{$\lambda_0$} [r] <-1pt,0pt> at -0.1  1
 \put{$\bullet$} [cc] at 1 1
  \endpicture   \]
\caption{Graph of the potential $v(z) $ for the two simplest  
choices of $\alpha$.}
\label{fig:pot} \end{figure}

 Consider the potential function 
 ($1 \le \alpha \in \mathbf{R}$; $\mathbf{r}=(x,y,z) \in 
\mathbf{R}^3$)   
 \begin{equation}v(\mathbf{r}) = \begin{cases} 0, & z<0 \cr
           \lambda_0 \left(\frac{z}{z_0}\right)^\alpha, &z>0. 
  \end{cases}
 \label{pot}\end{equation} 
It is characterized by $\alpha$ and the single length scale
 \[  \hat z 
 = \left(\frac{z_0{}\!^\alpha}{\lambda_0} \right)^{\frac1{\alpha+2}}.
\]
We ordinarily fix $\lambda_0 = z_0=1$ and let $\alpha$ 
vary (and suppress the arguments $x$ and~$y$). Thus
 $v(1)=1$ for all $\alpha$ and the potential forms an increasingly steep 
 wall near $z=1$ as $\alpha\to\infty$ (Fig.~\ref{fig:pot}).

 The scalar field is an operator-valued distribution 
 satisfying\footnote{We take $\hbar=1=c$, curvature coupling 
constant $\xi=\frac14$, and metric signature $g_{00}<0$.}
\begin{equation} 
   \frac{\partial^2\varphi}{\partial (x^0)^2} = 
 \nabla^2\varphi  - v(z)\varphi.
 \label{fieldeq}\end{equation}
Because the PDE (\ref{fieldeq}) is linear and invariant under 
translation in time~($x^0$), it has a standard reduction to a 
classical eigenvalue problem:  $\varphi(x^0,x,y,z)$ is a linear 
superposition (an integral, since the spectrum in our model will be 
continuous) of normal modes of the form
 $\phi_j(\mathbf{r}) e^{\pm i\omega_j x^0}$
 with operator coefficients.  For the details we refer to 
\cite{PMG,rect}.
The physical quantities of interest are 
 (the vacuum expectation values of) 
 the components of the energy-momentum tensor, especially the energy 
density
 \begin{equation}
 T^{00}(\mathbf{r})
  = \frac12\left[\left(\frac {\partial\varphi}{\partial x_0}\right)^2
  -\varphi\nabla^2\varphi+ v\varphi^2\right].
 \label{T00}\end{equation}

 These expectation values can be expressed as derivatives of the 
 cylinder kernel (Poisson kernel) 
 of the  differential operator on the 
 right-hand side of (\ref{fieldeq}), which can be defined by the PDE
 \begin{equation}
\left(\frac{\partial^2}{\partial t^2} + 
 \nabla^2 -v(z)\right)\T(t,x,y,z,z') =
 2\delta(t)\delta(x)\delta(y)\delta(z-z')
 \label{Tpde}\end{equation}
or by the eigenfunction expansion  
 \begin{equation}
 \T(t,\mathbf{r},\mathbf{r}') = 
- \int d\mu(j)\, \omega_j{}\!^{-1}
 \phi_j(\mathbf{r})\phi_j(\mathbf{r}')
  \, e^{-\omega_j t},
 \label{Teigen}\end{equation}
 where $\mu$ is a properly normalized spectral measure over the 
index set.
($\T$ is an analytic continuation via $t= i(x-x')^0$ of the Green 
function of the wave equation (\ref{fieldeq}).  In (\ref{Teigen}) 
we take advantage of the translational invariances of 
(\ref{fieldeq}) to set $x'=y'=0$ without loss of generality.)
Now (formally)
  \begin{equation}
 T^{00} = 
\left.  -\,\frac12\,
  \frac{\partial^2\T}{\partial t^2}\right|_{t=0, x=0=y, z'=z}\,,
 \label{T00fromT}\end{equation}
 and there are similar formulas (involving derivatives with 
respect to the spatial variables) for the three pressures,
  $P_z= T^{zz}$ etc.
 As usual, we must refer to the references for complete 
  explanations.

 In (\ref{T00fromT}) and similar formulas, the integral kernel is 
being evaluated at equal arguments (``on diagonal'' or ``in the 
concidence limit''\!, depending on whether one is in mathematics 
or physics literature).
In studying a new model it is common to look first at the diagonal 
value of $\T$ itself (without any derivatives), which has the 
physical interpretation of vacuum expectation value of the square 
of the field~$\varphi$.

 \section{Analytical solution}

 \subsection{Eigenfunctions}

 Solving (\ref{fieldeq}) by separation of variables is a standard 
exercise in quantum scattering theory.
 The formal index $j$ in (\ref{Teigen}) is a triple $(k_x,k_y,p)$ 
with $\mathbf{k}_\bot\equiv(k_x.k_y)\in \mathbf{R}^2$
  and $p\in(0,\infty)$.
 The frequency $\omega_j$ is the positive solution of
 $\omega^2= \mathbf{k}_\bot{}\!^2+p^2$.
 The eigenfunction is
 \begin{equation}
 \phi_j(\mathbf{r}) = (2\pi)^{-1}
 e^{i\mathbf{k}_\bot\cdot\mathbf{r}_\bot} \phi_p(z)
 \label{eigenfn}\end{equation}
 where
 \begin{equation}
  \left(-\,\frac{\partial^2}{\partial z^2} + v(z) -p^2\right)
  \phi_p(z) = 0.
 \label{eigeneqn}\end{equation}
 When $z<0$, it must take the form
 \begin{equation}
 \phi_p(z) =\textstyle{\sqrt{\frac2\pi}} \, \sin [pz-\delta(p)] 
 \label{eigenasymp}\end{equation}
 for some real phase shift $\delta(p)$
 (not to be confused with the Dirac $\delta$ in (\ref{Tpde}) and 
(\ref{diracint})). 
 The normalization factors in (\ref{eigenfn}) and (\ref{eigenasymp})
  guarantee that the spectral measure $\mu$ is 3-dimensional 
 Lebesgue measure.

 When $z>0$, $\phi$ is best expressed as
 \begin{equation}
  \phi_p(z) = C(p) P_\alpha\bigl( z/\hat z,
  (\hat z p)^2\bigr)
 \label{Pform}\end{equation}
 where the function $P_\alpha(z,E)$ is a solution of
 \begin{equation}
 \left(-\,\frac{d^2}{dz^2} + z^\alpha -E\right) P_\alpha(z,E) = 0
 \label{Peq}\end{equation}
 that vanishes as $z\to\infty$.
 For small, integer $\alpha$ the solutions are known as Airy 
functions and parabolic cylinder functions:
 \begin{equation}
 P_1(z,E) \propto \Ai (z-E), \qquad
  P_2(z, E) \propto D_{\frac12(E-1)} (\sqrt2\, z).
 \label{smallalpha}\end{equation}
 For a hard wall at $z_0$, we have
  $  P_\infty(z,E) \propto \sin[\sqrt E (z-z_0)]$.
Henceforth we take $\hat z = z_0 = 1$ so that $E = p^2$.

The solutions and their derivatives must match at $z=0$.
 Thereby $C$ and, more importantly, $\delta$ are
 determined:
 \begin{equation}
 \tan\bigl(\delta(p)\bigr) = - p\, \frac{P_\alpha(0,p^2)} 
{P'_\alpha(0,p^2)}\,,
 \label{tandelta}\end{equation}
 \begin{equation}
 C(p)^2 = \frac2\pi\, \frac1{P_\alpha(0,p^2)^2 
+ p^{-2} P'_\alpha(0,p^2)^2} \,.
 \label{Cformula}\end{equation}
 Even in the cases (\ref{smallalpha}) these formulas do not lend 
themselves to exact evaluation of the integrals for energy density 
and pressure, so further approximation or qualitative analysis is 
needed.

\subsection{Asymptotics}

 When $p=0$ the solution of (\ref{Peq}) is known 
  (a modified Bessel function): 
 \begin{equation}
   P_\alpha(z,0) = z^{1/2} K_{\frac1{\alpha+2}} \left(\frac2{\alpha+2}\, 
 z^{\frac{\alpha+2}2}\right).
 \label{P0}\end{equation}
 For small $p$ the solution can therefore be constructed as a
 perturbation expansion:
 \begin{equation}
   P_\alpha(z,E) = P_\alpha(z,0) + E P_\alpha^{(1)}(z) + \cdots.
 \label{pert}\end{equation}
 This process requires constructing the Green function (resolvent 
kernel) for the nonhomogeneous unperturbed equation and applying it 
iteratively.
 In this way we find that
 \begin{equation}
\delta(p) = p \bigl(\alpha +2\bigr)^{\frac2{\alpha+2}}\,
 \Gamma\left(\frac{\alpha+3}{\alpha+2}\right)
 \Gamma\left(\frac{\alpha+1}{\alpha+2}\right)^{-1} +O(p^3).
 \label{deltasmall}\end{equation}
 Fortunately, the Bessel-function integral needed to find the 
$O(p^3)$ term can be evaluated in closed form, but we do not report 
the result here. 

 At  large $p$ one can construct a WKB (semiclassical) approximation:
\begin{equation}
\phi_p(z) \sim [p^2-v(z)]^{-\frac14} 
\cos\left[\int_z^a\sqrt{p^2-v(\tilde z)}\,d\tilde z -\frac\pi4\right],
 \label{WKB}\end{equation}
where $a= p^{2/\alpha}$ is the turning point.
It follows that
\[
  \delta(p) \sim \int_0^a\sqrt{p^2-v(z)}\,dz + \frac\pi4
  \mod \pi .
\]
 Closer examination shows that the ``$\mathrm{mod}\ \pi$'' 
 can be ignored and the integral evaluated as a beta function:
 \begin{equation}
\delta(p)  =\frac1\alpha\, p^{1+2/\alpha} \,
\mathrm{B}\!\left(\frac32,\frac1\alpha\right)+\frac{\pi}{4}
+o(1).
 \label{deltalarge}\end{equation}

  In summary, we have for  $\alpha=1$ (the Airy function) 
 \begin{equation}
\delta(p)\sim \begin{cases} p\,
 3^{2/3}\Gamma(\frac43)/\Gamma(\frac23), &p \to0, \cr
\frac{2p^3}{3} + \frac\pi4\,, & p\to\infty,\end{cases}
 \label{airydelta}\end{equation}
and for $\alpha=2$ (the parabolic cylinder function)
 \begin{equation}
\delta(p)\sim \begin{cases} 2p\,
 \Gamma(\frac54)/\Gamma(\frac34), &
p \to0, \cr
\frac{\pi p^2}4 + \frac{\pi}{4}\,, & p\to\infty.
 \end{cases}
 \label{parabdelta}\end{equation}

We are interested in the power 
potential (\ref{pot}) only as a convenient model with suitable 
qualitative properties.
 Since the function  $\delta(p)$  completely encodes the influence 
of the potential on the field in the potential-free region,
 it is tempting to forget the potential and study the class of 
models parametrized by functions $\delta$ in a suitable class.
 The asymptotic relations (\ref{deltasmall}) and (\ref{deltalarge}) 
give some idea of what an allowed $\delta$ must look like,
 but otherwise the inverse problem of determining $v$, or even its 
basic qualitative properties, from a given $\delta$ is wide open,
 as far as we know.  Later we shall show evidence that slight 
changes in $\delta$ can produce unacceptable results.

 \subsection{The renormalized cylinder kernel}

To exploit the symmetry between $t$ and $\mathbf{r}_\bot$ in this 
problem, we introduce another layer of Fourier transformation into 
(\ref{Teigen}):    
\begin{equation}
\overline{T}(t,\mathbf{r}_\bot,z,z') =
 \frac1{(2\pi)^{3/2}} \int_{\mathbf{R}^3} d\omega \,d\mathbf{k}_\bot
 \int_0^\infty dp\, e^{i\omega t} e^{i\mathbf{k}_\bot\cdot 
\mathbf{r}_\bot} \phi_p(z) \hat{T} (\omega, \mathbf{k}_\bot,p),
 \label{TFour}\end{equation}
 \begin{equation}
  \hat{T}(\omega,\mathbf{k}_\bot,p) = \frac{-2}{(2\pi)^{3/2}}\,
 \frac{\phi_p(z')}{\omega^2 + k_\bot{}\!^2+p^2}\,, 
 \label{Ttransf}\end{equation}
where $\omega$ has now become an independent parameter.

 The integral over $\mathbf{R}^3$ can be done by standard methods, 
resulting in
 \begin{equation}
\overline{T}(t,\mathbf{r}_\bot,z,z') = 
 -\,\frac1{2\pi}\int_0^\infty dp\,
Y(s, p) \phi_p(z)\phi_p(z'),
 \label{Tcart}\end{equation}
 \[
Y(s,p) \equiv \frac{e^{-sp}}{s} \,, \qquad s \equiv
\sqrt{t^2+|\mathbf{r}_\bot|^2}\,.
 \]
We shall concentrate for now on the potential-free region, $z<0$,
where
 \begin{equation}
\overline{T} = -\,\frac1{\pi^2} \int_0^\infty dp\, Y(s,p) 
\sin\bigl(pz - \delta(p)\bigr) \sin \bigl(pz'-\delta(p)\bigr) .
 \label{Tcartouter}\end{equation}
Upon converting the product of sines to a sum of cosines in the 
standard way, 
one sees that the first term is just the ``free'' kernel that would 
apply in $\mathbf{R}^4$ if the potential were not there: 
 \begin{eqnarray}\label{freeren}
 \overline{T} &=&-\,\frac1{2\pi^2} \, 
\frac1{t^2+\mathbf{r}_\bot{}\!^2 +(z-z')^2}
 \\ && {}+
 \frac1{2\pi^2} \int_0^\infty 
dp\, Y(s,p) \cos\bigl(p(z +z')-2\delta(p)\bigr) 
\nonumber  \\ 
 &\equiv& \T_\mathrm{free} + \T_\mathrm{ren}\,.
 \nonumber\end{eqnarray}
 
    For a hard (Dirichlet)  
  wall at $z=z_0$ we have  $\delta(p)= z_0p $ 
 and hence the well known image solution,
 \begin{equation}
 \T_\mathrm{ren} = \frac1{2\pi^2} \, 
\frac1{t^2+\mathbf{r}_\bot{}\!^2 +(z+z'-2z_0)^2}\,.
 \label{image}\end{equation}
 Before continuing it is instructive to take a close look at this 
case.
 $\T_\mathrm{free}$ is, of course, singular on the diagonal 
 ($t=0$, $\mathbf{r}_\bot=0$, $z'=z$) and only there.
 The singularity makes it impossible to pass to the diagonal
 directly in (\ref{T00fromT}), but that is also unnecessary:
 $\T_\mathrm{free}$ is present in all problems, including empty 
space (where $T^{\mu\nu}$ is naturally defined to be zero), and 
hence is physically meaningless.  One expects to isolate and 
discard it before implementing (\ref{T00fromT}), which is applied 
only to the ``renormalized'' kernel, $\T_\mathrm{ren}\,$.
 The latter, as given in  (\ref{image}),
 is nonsingular in the region of physical interest; on diagonal one 
gets (with our convention $z_0=1$)
 \begin{equation}
 \T_\mathrm{ren} = \frac1{8\pi^2}\, \frac1{(z-1)^2}
 \label{wallTren}\end{equation}
 and similar formulas (proportional to $(z-1)^{-4}$) for the energy 
density and pressures.
 In the present case (\ref{Tcartouter}) is applicable over the whole 
range $-\infty< z < 1$, and (\ref{wallTren}) gives the 
expectation value of the scalar field right up to the wall
 (where it develops a nonintegrable divergence).
 This much is totally standard and familiar to all workers in the
field of vacuum energy.
However, let us go back to the integral form of 
(\ref{image})  contained in (\ref{freeren}):
 \begin{equation}
\T_\mathrm{ren} = \frac1{2\pi^2} \int_0^\infty 
dp\, \frac{e^{-sp}}{s} \cos\bigl(p(z +z')-2p\bigr) ,
 \label{imageint}\end{equation}
 where $s=\sqrt{t^2+|\mathbf{r}_\bot|^2}$,
 and attempt to set $t$ and $\mathbf{r}$ equal to 0 before 
evaluating the integral.  Although $\T_\mathrm{ren}$ is 
 well-defined except when $z+z'=2$, in (\ref{imageint}) we appear 
to have a double disaster: the denominator of the integrand is 
identically zero, and, moreover, even the integral of the numerator 
alone diverges because the exponential cutoff is lost.
 The resolution of this apparent paradox is that 
 \begin{equation}
\int_0^\infty \cos pz \, dp =\pi \delta(z) 
\label{diracint} \end{equation}
in the sense of distributions, and the Dirac distribution 
$\delta(z)$ is identically $0$ for $z\ne0$.
 Thus the classically divergent numerator integral is  equal 
to $0$ in the distributional sense in the limit $s\to0$,
 and  (\ref{imageint}) is consistent with (\ref{image}) and 
(\ref{wallTren}) there.

 Returning to the general case, we are confronted by the integral
 \begin{equation}
 \T_\mathrm{ren} = 
\frac1{2\pi^2} \int_0^\infty dp\, \frac{e^{-sp}}{s}\,
  \cos\bigl(p(z +z')-2\delta(p)\bigr).
 \label{Trencart}\end{equation}
 One might consider evaluating 
 it numerically,
  given a trustworthy formula or ansatz for $\delta(p)$.
  However, in view of the previous special example, it is not 
  surprising that the integral   
 is poorly convergent when $s\equiv\sqrt{t^2 +|\mathbf{r}_\bot|^2}$
  is small, 
which is precisely where we want it.  In fact, we should be able to 
take $s=0$ and get a finite answer when $z+z'>0$, but instead we 
have the same apparent infinities as in the Dirichlet case;
 and this time it is not obvious that the integral (without the 
factor $s^{-1}$) vanishes distributionally to lowest order in $s$ 
when $z+z'>0$, although that must surely be true. 

  To compound the problem, it appears that even this weak kind of
  convergence depends sensitively on $\delta$.
 Suppose that instead of (\ref{imageint}) we had considered 
\[
 \frac1{2\pi^2} \int_0^\infty 
dp\, \frac{e^{-sp}}{s} \cos\bigl(p(z +z')-2p
 +{\textstyle\frac{\pi}2}\bigr) ,
 \]
 which one might naively think to correspond to the large-$\alpha$ 
limit of  (\ref{deltalarge})
(which actually is invalid for $\alpha\to\infty$ with fixed~$p$).
 This integral equals
 \[
-\,\frac1{2\pi^2} \, \frac{z+z'-2}{\sqrt{t^2+\mathbf{r}_\bot{}\!^2}}
 \,\frac1{t^2+\mathbf{r}_\bot{}\!^2 + (z+z'-2)^2}\,.
\]
 Thus the cancellation that removes the divergence on the $z$ axis 
in (\ref{imageint}) does not happen here.
 More generally,
 there is a genuine divergence for $\delta(p) = Ap +B$ unless 
$B=0$.  
 Naively one would think that the divergent boundary energy we are 
studying is contributed by the modes of large~$p$, and that 
therefore 
 only the leading term in the WKB asymptotics (\ref{deltalarge})
  would be significant;
 the current example shows that that is not true.
 The fallacy in the reasoning is that high frequency can correspond 
to large $\mathbf{k}_\bot$ at fixed~$p$, as well as to large~$p$.

The sensitivity of (\ref{imageint}) to a constant phase shift 
remains visible in the polar framework treated in the next 
subsection.  The $u$ integral in (\ref{Tpolar}) or 
(\ref{Tpolardiag}) in that case 
 evaluates to a Bessel function $J_1\,$, which 
decays slowly as $\rho\to\infty$; however, the outer integration 
yields (\ref{wallTren}) by a standard formula found in handbooks 
and known to {\sl Mathematica}.  But the tiniest phase shift 
augments the Bessel function by a Struve function, which approaches 
a nonzero constant at infinity, so that the integral  diverges 
unambiguously.

 \subsection{Polar coordinates}

 Therefore, we recast the integration so that all high frequencies 
are treated on an equal footing.
 Any true divergence must come from the integral over high 
frequencies, since the eigenfunctions are smoooth and bounded.

 Abandoning the key formula (\ref{Trencart}) for now, we return to 
(\ref{TFour}), which in the notations
 $Z\equiv z+z'$, $\mathbf{s}=(t,\mathbf{r}_\bot)$, 
 $\mathbf{v}=(\omega,\mathbf{k}_\bot)$,
becomes
  \begin{equation}
  \T_\mathrm{ren} 
 =\frac1{4\pi^4}\int_0^\infty dp \int_{\mathbf{R}^3} d\mathbf{v} 
\frac{e^{i\mathbf{v}\cdot\mathbf{s}}}{v^2+p^2}\, 
 \cos\bigl(pZ-2\delta(p)\bigr).
 \label{Tvecfour}\end{equation}
 (Note that $s\equiv |\mathbf{s}|$ is the same $s$ as before.)
Now introduce polar coordinates in the  space of variables 
 $(p,v_1,v_2,v_3)$, with
 the main axis in the $Z$ direction and the prime meridian through 
$\mathbf{s}$ (that is, $s_2=0=s_3$).
 After several steps of calculation one arrives at the new key 
formula
 \begin{equation}
 \T_\mathrm{ren}  = \frac1{\pi^3} \int_0^\infty d\rho \int_0^1du\,
s^{-1} \sin(s\rho\sqrt{1-u^2})
  \cos\bigl(Z\rho u- 2\delta(\rho u)\bigr) .
 \label{Tpolar}\end{equation}
 Because of the sine, the integrand is not singular, although it 
needs to be defined by a limit when $s=0$.
 Ultimately we would like to take derivatives of (\ref{Tpolar}) and 
then pass to the diagonal, but for now we set
  $s=0$ and $z=z'$ immediately:
\begin{equation}
 \T_\mathrm{ren}(0,0,0,z,z) = \frac1{\pi^3} \int_0^\infty 
d\rho \int_0^1du\, \rho\sqrt{1-u^2} 
 \cos\bigl(2z\rho u- 2\delta(\rho u)\bigr).
\label{Tpolardiag} \end{equation}
 (As previously remarked, this should yield the expectation value of 
$\varphi^2$ at~$z$.) 

\begin{figure}
\scalebox{.6}{\includegraphics{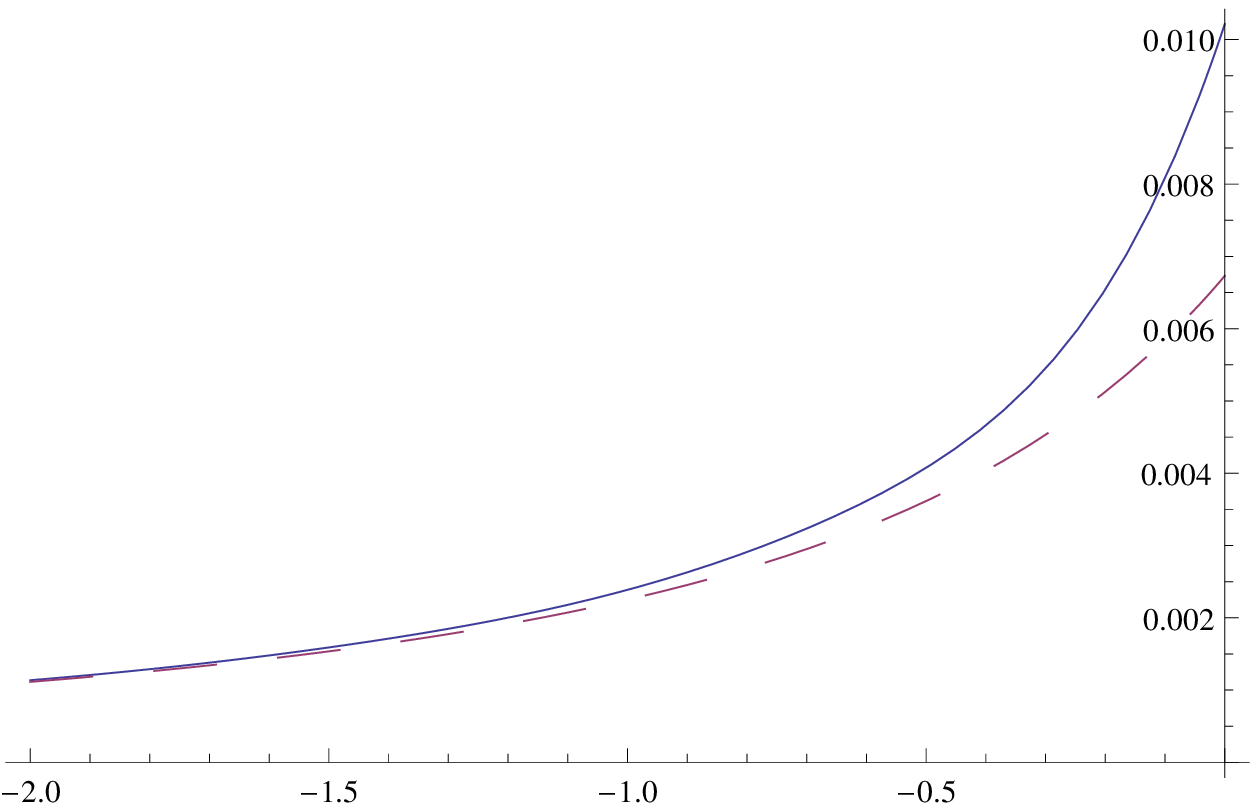}}
\caption{(Solid) $\T_\mathrm{ren}$ (\ref{Tpolardiag}) as a 
function of $z$ for $\alpha=1$.  
(Dashed) $\T_\mathrm{ren}$ for a hard wall at $ 
c=3^{2/3}\Gamma(\frac43)/\Gamma(\frac23)\approx 1.37172$.}
\label{fig:tgraph}\end{figure}

 The change of variables has not eliminated the problem, but it has 
isolated it at the upper limit of a single improper integral.
 The convergence of (\ref{Tpolardiag}) is still delicate.
 Numerical and analytical investigations of it are ongoing.
 Were it not for the convergence issues, 
one could prove easily from (\ref{Tpolardiag}) that the function
manifests approximately 
inverse-square decay  resembling  (\ref{wallTren}), but with 
the efffective wall position $z=1$ replaced by $z=c$, where $c$
is the coefficient of the linear term in (\ref{deltasmall}).
Numerical integrations have been performed in {\sl Mathematica}
for $\alpha=1$.
Despite the instability of the highly oscillatory integrals,
the results are qualitatively as expected, approaching 
$(z-c)^{-2}/8\pi^2$ already for moderately large $|z|$ 
(Fig.~\ref{fig:tgraph}).

As this paper was being drafted, K. Milton et al.\ reported 
(privately; cf.\ \cite{miltonpaper}) an independent evaluation of 
$\T_\mathrm{ren}$ 
for $\alpha=1$ using 
a different integral representation with better convergence 
properties.  They find a weak ($z^{-1}$) divergence in the 
energy density at the origin, which in hindsight is to be 
expected as diffraction from the coefficient singularity there 
(which will become less important for larger~$\alpha$).  The 
corresponding singularity in $\T$ is of order $z \ln|z|$, 
therefore not visible in Fig.~\ref{fig:tgraph}.

 \subsection{Conclusions}

 Our principal results are the Cartesian formula (\ref{Trencart}),
 the polar formula (\ref{Tpolar}), and the latter's diagonal 
specialization (\ref{Tpolardiag}),
 along with the formula (\ref{tandelta}) for $\delta$ and the 
asymptotic analyses in the subsection following it. 
 Formulas for energy density and pressure can be obtained by 
differentiating (\ref{Tpolar}) and again passing to the diagonal.  
Granted the convergence of the integrals, it can be seen 
 that $T^{00}$ and $T^{11}$ (the pressure in the $x$ direction) are 
equal and opposite, so that (\ref{enbalance}) is satisfied in the 
sideways motion of a plane partition perpendicular to the plane 
wall.  This happy result, which fulfills the main motivation of the 
 project,   follows from the finiteness of the 
diagonal values (in the potential-free region) of the kernel 
$\T_\mathrm{ren}$ and its derivatives without the need of an 
artificial cutoff, since it is satisfied by the individual normal 
modes in the integrand.


 \subsection{Inside the wall}

 Detailed calculations in the region $z>0$ have not yet been 
attempted, since we want to exploit fully the more elementary 
calculations in $z<0$ first.  Also, one may reasonably consider the 
energy and stress in that region to be part of the wall, not part 
of the cavity containing the field.  Nevertheless, in our setup the 
properly renormalized stress tensor inside the wall should still be 
finite and display the physically sensible behavior (\ref{enbalance}).
 (Most of the arguments of the previous subsection still apply 
here, or can be adapted.)

 On general grounds one knows that the renormalization will require 
subtraction of additional terms from $\T$.  In a nonsingular model 
the diagonal divergences in $\T$ and hence $T^{\mu\nu}$ are 
determined \cite{systematics} by the well known small-time divergences 
of the heat (or quantum) kernel \cite{gilkey,wilk}.
 Wherever the potential $v(\mathbf{r})$ is not zero, the energy 
density calculated with the standard ultraviolet cutoff ($t$~held 
nonzero) has the expansion
 \begin{equation}
\pi^2 T^{00} \sim {\textstyle\frac32} t^{-4} 
 -{\textstyle\frac18} v t^{-2} 
 +{\textstyle\frac1{32}}(v^2 -
 {\textstyle\frac13}\nabla^2v)\ln t
\quad \mbox{as $t\to0$}.
 \label{vdivs}\end{equation}
 The first term corresponds to the universal vacuum energy that has 
been removed by subtracting the free kernel at (\ref{freeren}).
 The  other terms in (\ref{vdivs}) can be removed by subtracting 
higher-order terms in the small-$t$ expansion of the  exact $\T$ 
(obeying (\ref{Tpde}) with the potential included).  Physically, 
such terms represent redefinitions of the constants in the equation 
of motion of the $v$ field itself.

 A technical issue that must ultimately be addressed here is the 
validity of (\ref{vdivs}) when $v\notin C^{\infty}$.  In 
particular, for $\alpha=1$ (\ref{vdivs}) predicts a Dirac delta at 
$z=0$ from the term $\nabla^2v$,
  and we will not trust the numerical  coefficient until a direct 
  calculation has been carried out.
 The issue here is the same one that arises in the heat kernel (and 
Gauss--Bonnet theorem) for the Laplacian in a region in the plane:
 The contributions of the corners of a polygon cannot be obtained 
by naively taking the limit of the effects of the curvature of a 
smooth boundary.  This is an interesting question to be 
investigated in the future.

 \section{Semiclassical analysis}

 \subsection{General remarks}

 Separation of variables, even when it is available, is not always 
the best way to obtain information about the integral kernels (and 
spectral measures)
associated with a linear partial differential operator.
Leading terms such as those in   (\ref{vdivs}) are routinely found 
by direct construction of some kernel as an asymptotic series.
Higher-order information can be obtained from terms in the quantum 
kernel (Green function of the time-dependent Schr\"odinger 
equation) corresponding to periodic orbits of the 
underlying classical mechanical system 
\cite{BB3,BB5,gutz,zelditch}.
 (These terms are also present in the heat kernel but exponentially 
suppressed.  They produce oscillatory terms in the averaged 
eigenvalue density.)

 The construction of Green functions for the Laplace and Helmholtz 
equations in bounded domains in $\mathbf{R}^n$ (billiards) 
 by reduction to 
integral equations on the boundary is well known.  The counterpart 
construction for the heat equation is less familiar but available 
in the literature \cite{kress,rubinstein}.
 What is seldom appreciated is that for the heat equation the 
solution of the boundary integral equation by iteration is 
convergent, because the integral operator has Volterra structure. 
One therefore has, in principle, an explicit construction of the 
solution.  The Schr\"odinger equation has the same Volterra 
structure, so one expects again to have a convergent series 
solution. 
   To implement this idea in a general context, Mera 
\cite{merapaper}  has proved the following 
 \emph{general Volterra theorem\/}:

 \begin{theorem}
 Let the kernel $A(t,\tau)$ be 
 (for each $t$ and $\tau$ in an interval~$I$)
 a uniformly bounded linear  
operator  $A\colon{\mathcal B}\to{\mathcal B}$, where ${\mathcal B}$ is 
a Banach space, and  suppose that it has the Volterra property,
 $A(t,\tau)=0$ when $\tau>t$.
 Define the integral operator 
 $Q \colon L^\infty(I;{\mathcal B}) \to L^\infty(I;{\mathcal B})$ by
 \begin{equation}
Q\phi(t) = \int_0^t A(t,\tau) \phi(\tau)\, d\tau.
 \label{Voltker}\end{equation}
Then the Volterra integral equation 
\begin{equation}
 \phi - Q\phi = f \quad (f\in L^\infty(I;{\mathcal B}))
\label{Voltinteq} \end{equation}
 can be solved by successive approximations.
 That is, the Neumann series converges in the topology of 
$L^\infty(I;{\mathcal B})$.
\end{theorem}

 The application of the theorem in any particular context reduces 
to showing that the operator family $A(t,\tau)$ is uniformly 
bounded on a suitable space~$\mathcal B$.  For Schr\"odinger 
equations this is a nontrivial task and requires supplementary 
technical assumptions.
 Here we are primarily interested
 in problems with potentials in $\mathbf{R}^n$.
 In that setting the key idea, due to Balian and Bloch \cite{BB5},
  is to 
let the semiclassical or WKB approximation to the quantum kernel 
play the role played  by the free kernel in billiard 
problems, so that the role played by scattering off the boundary in 
billiards (or by scattering by the potential in standard 
 time-dependent perturbation theory 
\cite[Ch.~7]{merathesis}\cite{merapaper}) is played 
here by scattering by a source that is essentially the residual 
error in the WKB approximation to the exact kernel.
 This construction is developed in \cite[Ch.~8]{merathesis}.

 The WKB kernel is\footnote{In this section $\mathbf 
x$ and $\mathbf y$ are two different spatial points, not 
coordinates of the same point as earlier, and we reintroduce 
$\hbar$ to make the structure of the semiclassical asymptotics 
clearer.  To simplify the Schr\"odinger equation we take the mass 
$m=\frac12$.}
 \begin{equation}
G_\mathrm{scl}(\mathbf{x},t;\mathbf{y},0) = 
 (2\pi i\hbar)^{-n/2}\, A\, e^{iS/\hbar},
 \label{WKBker}\end{equation}
 where 
 \begin{equation}
 S(\mathbf{x},\mathbf{y},t) = \int_0^t L\bigl(\mathbf{q}(\tau),\dot 
{\mathbf q}(\tau)\bigr)\,d\tau,        
 \qquad L = {\textstyle \frac14} \dot{ \mathbf q}^2 - v(\mathbf{q}),
 \label{action}\end{equation}
is the classical action, a solution of the Hamilton--Jacobi 
equation, and the \emph{amplitude\/} $A$ is
 \begin{equation}
A(\mathbf{x},\mathbf{y},t)= 
  \sqrt{\det 
 \left(-\,\frac{\partial^2S}{\partial x_i\,\partial x_j}\right)} 
\,. 
 \label{amp}\end{equation}
 If there is more than one classical 
trajectory $\mathbf{q}(\tau)$ starting at $\mathbf y$ at time $0$ 
 and arriving at $\mathbf x$ at time~$t$, the semiclassical 
approximation is a sum of such terms, possibly modified by Maslov 
phase factors (see next subsection)  to keep track of places where 
the radicand in (\ref{amp}) has gone negative.

 Define a kernel $Q$ by 
 \begin{equation}
Q(\mathbf{x},t;\mathbf{y},\tau) = 
\hbar^2 [\Delta_\mathbf{x} A(\mathbf{x},t;\mathbf{y},\tau)]
  e^{iS(\mathbf{x},t;\mathbf{y},\tau)}.
 \label{Qdef}\end{equation}
The corresponding operators  $Q$  and $G_\mathrm{scl}$ are related 
by
 \begin{equation}
\bigl(-i\hbar \partial_t - \hbar^2\nabla^2 +v(\mathbf{x})\bigr)
 G_\mathrm{scl} = 1 - Q\,;
 \label{Qeq}\end{equation} 
that is, $Q=O(\hbar^2)$ 
 is the amount by which $G_\mathrm{scl}$ fails to solve 
the PDE for which it was devised.
 Thus, formally,
 \begin{equation}
 G = G_\mathrm{scl} \sum_{j=0}^\infty   Q^j,
\label{Gopseries}\end{equation}
 or
 \begin{equation}
G(\mathbf{x},t;\mathbf{y},\tau) =G_\mathrm{scl}(\mathbf{x},t;\mathbf{y},\tau) 
 + \int_0^t \Gamma(t,\tau_1)\Lambda(\tau_1,\tau)\, d\tau_1 + \cdots,
 \label{Gkerseries}\end{equation}
 where 
 \begin{equation}
 [\Gamma(t,\tau)\phi(\tau)](\mathbf{x}) = \int_{\mathbf{R}^n} 
G_\mathrm{scl}(\mathbf{x},t;\mathbf{y},\tau) \phi(\mathbf{y},\tau)\, 
d\mathbf{y},
\label{Gammadef}\end{equation}
 \begin{equation}
 [\Lambda(t,\tau)\phi(\tau)](\mathbf{x}) = \int_{\mathbf{R}^n} 
Q(\mathbf{x},t;\mathbf{y},\tau) \phi(\mathbf{y},\tau)\, 
d\mathbf{y}.
\label{Lambdadef}\end{equation}
                                              
 \begin{theorem} \cite{merathesis}
 \label{thm:WKB}
 In the notation of the two foregoing paragraphs:
  Suppose that the following two hypotheses hold:
 \begin{enumerate}
 \item $ \left\|\frac{\Delta A}
 A\right\|_{L^\infty(I^2;\mathbf{R}^{2n})}<\infty$.
 \item  $\Gamma$ is a bounded operator from $L^2(\mathbf{R}^n)$ to 
itself.
 \end{enumerate}
 Then the semiclassical operator
 $\Lambda\colon L^2(\mathbf{R}^n) \to L^2(\mathbf{R}^n)$ is a 
bounded linear integral operator.
 It follows that the Volterra integral equation in the space 
$L^{\infty.2}(I;\mathbf{R}^n)$ with the semiclassical kernel
 $Q(\mathbf{x},t;\mathbf{y},\tau)$ can be solved by successive
 approximations.
 \end{theorem}

 This construction implements the Feynman \emph{path integral} idea 
in a way different from the usual time-slicing approach.
 Each term in (\ref{Gkerseries}) is an integral over classical 
paths with $j$ scatterings off an effective potential
 $\Delta A/A$. 
 
 The determinant in 
(\ref{amp}) is singular at \emph{caustics}, where the mapping 
from initial velocity data (at~$\mathbf y$) to $\mathbf x$ ceases 
to be a diffeomorphism. 
 One can expect both conditions (1) and (2) to be 
problematical if the orbit goes through a caustic,
 but we provide some evidence below that the situation is not as 
bad as one might expect.
A way  to go beyond caustics (if necessary)
 is provided by the Maslov theory \cite{MF}, 
as already implemented in a similar problem in \cite{zapata}.

 \subsection{The harmonic oscillator and the quadratic wall}

 It was natural to apply Theorem \ref{thm:WKB} to a power 
potential, with two motivations:  
 to test the validity of the two hypotheses
in the theorem in the context of a concrete problem, and to seek 
new information about the spectral density (and hence eventually 
the vacuum energy) for a soft wall.  We have studied the case 
$\alpha=2$ in one dimension,
 \begin{equation}
v(x) = \begin{cases}
             0 &\mbox{if $x\le0$}, \\
            \frac14 \omega^2 x^2 & \mbox{if $x>0$}.
 \end{cases}
 \label{harmwall}\end{equation}
 The two transverse dimensions can be ignored because their 
contribution to the quantum kernel in dimension~$3$ is a trivial 
factor.

 \begin{figure}
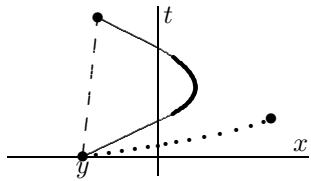

\[\beginpicture
 \setcoordinatesystem  units <1cm, 0.5cm>
 \putrule from -2 0 to 2 0
 \putrule from 0 -0.5 to 0 4
 \put{$\bullet$} at -1 0
  \put{$\bullet$} at -0.8  3.7
 \put{$\bullet$} at 1.5 1
 \plot -1 0
  0 0.95 /
 \plot 0 2.9  
       -0.8 3.7 /
 \setdashes\noindent\plot -1 0 
  -0.8 3.7 /
 \setquadratic
\setsolid\noindent\plot 0 0.95
 0.5 1.85
 0 2.9 /  
 \setplotsymbol({\bf.})
\noindent\plot  0.2 1.15
 0.5 1.85
 0.2 2.65 /
  \setdots\noindent\plot 0 0.3
 1 0.7
 1.5 1 /
\setlinear\noindent\plot -1 0
       0 0.3 /
\put{$x$} [rb] <0pt, 2pt> at 2   0
\put{$t$} [lt] <2pt, 0pt> at 0 4
\put{$y$} [t] <0pt, -2pt> at -1 0
\endpicture\]
 \caption{Paths with actions and amplitudes calculated in the text.
 Dashed: (5.13). Solid: (5.15) (heavy: (5.16--17)).  Dotted:
 (5.19--20).}
 \label{orbits}\end{figure}

Consider points $x$ and $y$ that are both in the 
potential-free region, and a time $t>0$. 
  There is always a force-free motion 
directly from $y$ to $ x$ in time~$t$.  
 It is elementary to calculate  for this direct path
\begin{equation}
S = \frac{(x-y)^2}{4t}\,, \qquad A^2 = \frac1{2t}\,, 
 \qquad \Delta A =0,
 \label{direct}\end{equation}
 so that (of course)  the quantum kernel (\ref{WKBker})
  is just that of a free particle.

  If $\omega t>\pi$ there is another classical path that enters the 
harmonic-oscillator region at
 \begin{equation}
 \tau = t_1 =  \frac y{x+y}\left(t-\frac{\pi}{\omega}\right),
 \label{t1-0refl}\end{equation}
 and reemerges after half a period,
at $t_2 =t_1 + \pi/\omega$.  
 This path also contributes to the 
leading term, $G_\mathrm{scl}\,$, in (\ref{Gkerseries}).
 (Bear in mind that such contributions are always of the schematic 
form $Ae^{iS}$, which we shall not constantly repeat.)
  We shall see that the action of the 
portion of the orbit inside the potential is~$0$, so the total 
action is just that of the two free motions at the ends:
 \begin{equation}
S = \frac{y^2}{4t_1} + \frac{x^2}{4(t-t_2)}=
  \frac{(x+y)^2}{4\left(t-\frac{\pi}{\omega}\right)}\,,
 \qquad A^2 =-\, \frac1{2\left(t-\frac{\pi}{\omega}\right)}\,,
\qquad \Delta A =0.
 \label{0refl}\end{equation}
 Note that the resulting term 
  added to $G_\mathrm{scl}$  differs by a time translation 
 (and a phase, since $A^2$ is negative)
from the image term that would be produced by a hard wall.
We shall show that the proper
    phase factor is~$-i$.
  There is an apparent 
singularity in (\ref{0refl}) at $\omega t=\pi$ that deserves closer 
examination.

 Theorem \ref{thm:WKB} is formulated in 
\cite{merathesis} for a $C^\infty$ potential. 
For (\ref{pot}), in addition to (\ref{0refl}) there are
waves diffracted from the coefficient singularity at $z=0$, but 
they become increasingly negligible with increasing~$\alpha$.

 Now consider $x$ and $y$ both inside the potential.  From 
(\ref{action}) and the relevant solution of the classical equation 
of motion one can reproduce  well known formulas, 
 \begin{equation}
S(x,y,t) = \frac{\omega}{4\sin(\omega t)}\, 
 [(x^2+y^2)\cos(\omega t) -2 xy],
 \label{HOaction}\end{equation}
 \begin{equation}
 A^2 = \frac{\omega}{2\sin(\omega t)}\,, \qquad \Delta A = 0.
\label{HOamp}\end{equation}
 We need these formulas only for $0<\omega t< \pi$;
 however, for  the full harmonic oscillator potential on the whole 
real line it is well known \cite{MF,TT} that the resulting 
 (Mehler) formula 
for $G_\mathrm{scl}$ remains valid everywhere in space-time and 
gives the \emph{exact} quantum kernel, with the caveat that (as 
suggested by the sign change in (\ref{HOamp})) the kernel must be 
multiplied by $(-i)^\mu$ where $\mu$ is the number of occasions when
  $t$ has passed through  an integer multiple of $\pi/\omega$.
 (Conventionally  one redefines $A^2$ and $A$ to be always positive 
numbers and writes the Maslov phase factor $(-i)^\mu$ separately.)
At such a time there is a caustic;
 all the trajectories from $y$ refocus at 
 $x=(-1)^\mu y\,$.
But the kernel formula (\ref{WKBker}) reproduces there the original 
($t=0$) singularity, which is still a solution of the homogeneous 
Schr\"odinger equation.  (This situation is strikingly different 
from that for elliptic equations, such as the time-independent 
 Schr\"odinger equation, where a caustic marks the \emph{breakdown} 
of the semiclassical approximation.  Note that the celebrated 
\emph{turning points} of the harmonic oscillator are \emph{not} 
caustics for the time-dependent problem!)

 With this background understanding we can finish treating the 
 trajectory (\ref{0refl}):
 \begin{itemize}
 \item A variant of the calculation leading to (\ref{HOaction}) 
shows that, as claimed, $S=0$ for any trajectory linking $y=0$ to 
$x=0$ (necessarily in elapsed time $\pi/\omega$).
 \item As for the harmonic-oscillator kernel, the singularity in 
$A$ of (\ref{0refl}) is harmless; the companion factor 
$e^{iS/\hbar}$ is effectively $0$ there.
 \item By continuity from the (purely harmonic) case $y=0$, when 
$y<0$ but small one would expect a caustic to occur somewhere near
 $x=-y$, $t=\pi/\omega$.  Therefore, when the trajectory reemerges 
from the potential, this term of the kernel should be multiplied by 
a Maslov factor~$-i$.
\end{itemize}\goodbreak

To verify this last claim, and to make a start on computing the 
second (single-reflection) term in (\ref{Gkerseries}), 
 we consider a path that starts at $y<0$ at time $0$ and ends at 
$x>0$ at time $t$.
 It must cross the time axis at a time $t_1\,$, and from the 
solution of the classical equation one finds
 \begin{equation}
 \omega xt_1 + y\sin\bigl(\omega(t-t_1)\bigr) =0,
 \label{t1eq}\end{equation}
which can't be solved by elementary functions.
 The action is
 \begin{equation}
 S(x,y,t)=\frac{y^2}{4t_1} + \frac{y^2}{8\omega t_1{}\!^2}
 \sin\bigl(2\omega(t-t_1)\bigr).
  \label{ScaseV}\end{equation}
 By implicit differentiation of (\ref{t1eq}) one can find that
 \begin{equation}
A^2 = \frac{y}{2t_1}\,\frac1{x-y\cos(\omega(t-t_1))}\,.
 \label{AcaseV}\end{equation}
There will be a caustic if the denominator of (\ref{AcaseV}) 
changes sign.  (Since that factor arises from 
 $\partial t_1/\partial x$, its vanishing says that 
 $t_1$ (hence~$y$) can vary
without changing $x$ (at least to first  order).)

\begin{figure}
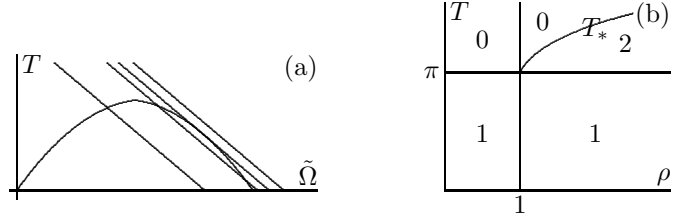

\[ \beginpicture
\setcoordinatesystem units <1cm,1.2cm>
\putrule from -0.1 0 to 4 0
\putrule from 0 -0.1 to 0 1.5
 \put{(a)} [rt] at 4 1.5
 \put{$\tilde\Omega$} [rb] <0pt,2pt> at 4 0
 \put{$T$} [lt] <2pt,0pt> at 0 1.5
 \plot   1.355 1.414
         2.355 0.707
         3.355 0 / 
 \plot   1.2 1.414
         2.2 0.707
         3.2 0 / 
 \plot   1.55 1.414
         2.55 0.707
         3.55 0 / 
 \plot   0.5 1.414
         1.5 0.707
         2.5 0 / 
  \setquadratic
 \plot 0 0
 0.785 0.707
 1.57  1
 2.355 0.707
 3.14  0 /
 \endpicture \qquad\qquad
 \beginpicture
\setcoordinatesystem units <1cm,0.5cm>
\putrule from 0 0 to 3 0
\putrule from 0 0 to 0 5
 \putrule from 0 3.14 to 3 3.14
 \putrule from 1 0 to 1 5
 \put{(b)} [rt] at 3 5
 \put{$\rho$} [rb] <0pt,2pt> at 3 0
\put{$T$} [lt] <2pt,0pt> at 0 5
 \put{$0$} at 0.5 4
  \put{$0$} at 1.3 4.5
 \put{$2$} at 2.4 3.9
 \put{$1$} at 0.5 1.5
\put{$1$} at  2.0 1.5
 \setquadratic
 \plot 1 3.14
 1.5 4
 2.5 4.7 /
 \put{$\pi$} [r] <-2pt,0pt> at 0 3.14
 \put{$1$} [t] <0pt, -2pt> at 1 0
 \put{$T_*$} at 2 4.3
 \endpicture\]
 \caption{(a) The four possible relations between a diagonal line and 
the principal arc of the sine curve.
 (b) Resulting division of the parameter plane,
  labeled by intersection numbers.}
 \label{fig:intersects} \end{figure}

 To investigate further it is helpful to introduce dimensionless 
variables
 \begin{equation}
T= \omega t, \quad \Omega = \omega t_1\,, \quad 
 0< \tilde\Omega = T-\Omega <\pi, \quad \rho = -\,\frac yx >0.
 \label{dimless}\end{equation}
 Then (\ref{t1eq}) is
 \begin{equation}
0 = \rho \sin(\tilde\Omega) + \tilde\Omega 
 - T \equiv f(\tilde\Omega)
 \qquad (0<\tilde\Omega<\pi),
 \label{f}\end{equation}
which can be investigated graphically as the intersection of a 
trigonometric graph and a straight line.
 The number of intersections  can be 0, 2, or~1 
(Fig~\ref{fig:intersects}).
 There is one solution if $T<\pi$ 
 (that is, the straight line hits the axis below the sine curve).  
 There are no solutions if 
$\rho\le1$ and $T>\pi$, or if $\rho>1$ and $T> T_*\,$, where
 \begin{equation}
 T_* = \sqrt{\rho^2-1} +\cos^{-1} \left(-\,\frac1\rho\right).
 \label{Tstar}\end{equation}
There are two solutions if $\rho >1$ and $\pi\le T< T_*\,$.
 Finally, there is one solution if the straight line is tangent to 
the sine curve --- that is, $f'(\tilde\Omega)$ and 
$f(\tilde\Omega)$ are zero simultaneously --- which happens when 
 $\rho\ge1$ and $T=T_*\,$.
Furthermore,
 \begin{equation}
 0 = f'(\tilde\Omega) =\rho\cos(\tilde\Omega) +1
 \label{fprime}\end{equation} 
is the condition for the vanishing of the denominator of 
(\ref{AcaseV}).

 Now consider a fixed trajectory with a moving endpoint
 (that is, fix $y$ and $t_1$ and let $x$ and $t$ vary).
 When $t\approx t_1\,$, $f'(\tilde\Omega)$ is large and positive
 ($\rho\to +\infty$, $\cos(\tilde\Omega)\to 1$).
 Near the exit point, $t\approx t_2\,$,  
 $f'(\tilde\Omega)$ is large and negative
  ($\rho\to +\infty$, $\cos(\tilde\Omega)\to -1$).
 Therefore, every trajectory does pass through a solution of 
(\ref{fprime}) somewhere on its retreat from the potential.

 In future work we hope to complete the calculation of 
$\Delta A$ for trajectories with an endpoint inside the potential.
That will enable one to study whether the two conditions in 
Theorem \ref{thm:WKB} are satisfied in spite of the caustic, as 
they are for the Mehler kernel.
 If so, then one can tackle the second (single-reflection) term in
 (\ref{Gkerseries}) by concatenating a trajectory of the sort just 
studied with one of the time-reversed type.  For given $(x,y,t)$ 
outside the potential,
 one must integrate over all $(q,\tau)$ inside the potential for 
which such a trajectory exists.  From the taxonomy of paths given 
above, it is clear that as many as four trajectories can exist, so 
the term $G_\mathrm{scl}Q^1 = \int \Gamma\Lambda$ is a sum of four 
terms, each with a domain of integration that is a nontrivial 
subset of the region $0<q<\infty$, $0<\tau<t$.
(Negative $q$ do not contribute, because we saw earlier 
 ((\ref{direct}) and (\ref{0refl})) that 
$\Delta A=0$ there.)

 \section{Conclusion}

 The seemingly elementary model of a ``power wall'' has run into 
several rather profound mathematical problems that are worthy of 
mathematicians' attention.
 \begin{enumerate}
\item What phase shifts $\delta(p)$ correspond to potentials (or 
even nonlocal dynamics) qualitatively worthy of being called ``soft 
walls''?
 \item What phase shifts do and don't lead to finite oscillatory 
integrals  (\ref{Trencart}), (\ref{Tvecfour}),  (\ref{Tpolar}),
 (\ref{Tpolardiag}),   etc.?
 When the convergence is unstable, can numerical methods 
nevertheless be applied to such integrals?  Can they be 
analytically recast into more rapidly convergent integrals?
 \item What can one say in general about caustics in the 
semiclassical solution of time-dependent Schr\"odinger equations?
 Are they generically as harmless as in the Mehler formula, or as 
harmful as in elliptic problems?
 \end{enumerate}

 \section*{Acknowledgments}

 We thank Gabriel Barton,   Lev Kaplan, and Kim Milton
  for valuable remarks.

\end{document}